\documentclass[a4paper]{jpconf}
\usepackage{graphicx,amsmath}

\usepackage{hyperref,color}

\newcommand{\half}{{\frac{1}{2}}}
\newcommand{\mbf}[1]{\mathbf{#1}}
\renewcommand{\bar}[1]{\overline{#1}}

  \def\navyblue{\color[rgb]{0,0,0.6}}

\begin{document}
\title{Light-Front Quantization and AdS/QCD: \\An Overview}

\author{Guy F. de T\'eramond$^1$ and Stanley J. Brodsky$^2$}

\address{$^1$ Universidad de Costa Rica, San Jos\'e, Costa Rica}
\address{$^2$ SLAC National Accelerator Laboratory,
Stanford University, Stanford, CA 94309}


\begin{abstract}
We give an overview of the light-front holographic approach to strongly coupled QCD, whereby a confining gauge theory, quantized on the light front, is mapped to a higher-dimensional anti de Sitter (AdS) space. The framework is guided by the AdS/CFT correspondence incorporating a  gravitational background asymptotic to
AdS space which encodes the salient properties of QCD, such as the ultraviolet conformal limit at the AdS boundary at $z \to 0$, as well as  modifications of the geometry in the large $z$ infrared region to describe confinement and linear Regge behavior.  There are two equivalent procedures for deriving the AdS/QCD equations of motion: one can start from the Hamiltonian equation of motion in physical space time by studying the off-shell dynamics of the bound state wavefunctions as a function of the invariant mass of the constituents. To a first semiclassical approximation, where quantum loops and quark masses are not included, this leads to a light-front Hamiltonian equation which describes the bound state dynamics of light hadrons  in terms of an invariant impact variable $\zeta$ which measures the separation of the partons within the hadron at equal light-front time. Alternatively, one can start from the gravity side by studying the propagation of  hadronic modes in a fixed effective gravitational background. Both approaches are equivalent in the semiclassical approximation.  This allows us to identify the holographic variable $z$ in AdS space with the impact variable $\zeta$.  Light-front holography  thus allows a precise mapping of transition amplitudes from AdS to physical space-time. The internal structure of hadrons is explicitly introduced and the angular momentum of the constituents plays a key role.

\end{abstract}

\section{Introduction}

The AdS/CFT correspondence between gravity or string theory on a higher dimensional anti de Sitter (AdS) space and conformal field theories in physical space
time~\cite{Maldacena:1997re}
 has led to a semiclassical approximation for strongly-coupled QCD, which provides physical insights into its non-perturbative dynamics.
 The correspondence is holographic in the sense that it determines a duality between  theories in different number of space-time dimensions.
In practice, the duality provides an effective gravity description in a ($d+1$)-dimensional AdS
space-time in terms of a flat $d$-dimensional conformally-invariant quantum field theory defined at the AdS
asymptotic boundary~\cite{Gubser:1998bc, Witten:1998qj}
\begin{equation}
  e^{i W_{\rm CFT} [J=\Phi_0] } = Z_{\rm grav}[\Phi] =  \int \mathcal{D}  e^{i S_{eff}[\Phi]},
\end{equation}
where $W_{\rm CFT}$ is the generating functional of the conformal theory and $Z$ is the string partition function given in terms of fields $\Phi$ in AdS space.
Thus, in principle, one can compute physical observables in a strongly coupled gauge theory  in terms of a classical gravity theory.

Anti-de Sitter $AdS_{5}$ space is the maximally symmetric space-time
with negative curvature and a four-dimensional space-time boundary.
The most general group of transformations that leave the $AdS_{d+1}$ differential line element
\begin{equation} \label{eq:AdSz}
ds^2 = \frac{R^2}{z^2} \left(\eta_{\mu \nu} dx^\mu dx^\nu - dz^2\right),
\end{equation}
invariant, the isometry group, has dimensions $(d+1)(d+2)/2$. Five-dimensional Anti-de Sitter space AdS$_5$ has 15 isometries, in agreement with the number of generators of the conformal group in four dimensions.

Since the AdS metric (\ref{eq:AdSz}) is invariant under a dilatation of all coordinates $x^\mu  \to \lambda x^\mu$  and $z \to \lambda z$, it follows that the additional dimension, the holographic variable $z$, acts as a scaling variable in Minkowski space: different values of $z$ correspond to
different energy scales at which the hadron is examined.  As a result,
a short space-like or time-like invariant interval
near the light-cone, $x_\mu x^\mu \to 0$,  maps to the conformal  AdS boundary near $z \to 0$. This also corresponds to the
$Q \to \infty$ ultraviolet (UV) zero separation distance. On the other hand, a large invariant four-dimensional  interval of
confinement dimensions $x_\mu x^\mu \sim 1/\Lambda_{\rm QCD}^2$ maps to the large infrared (IR) region
of AdS space $z \sim 1 / \Lambda_{\rm QCD}$.

QCD is fundamentally different from conformal theories since its scale invariance is broken  by quantum effects.
A gravity dual to QCD is not known, but the mechanisms of confinement can be incorporated in the gauge/gravity correspondence by modifying the AdS geometry
near a large infrared value  $z \sim 1/\Lambda_{\rm QCD}$, where $ \Lambda_{\rm QCD}$, sets the scale of the strong interactions. In this simplified  approach, hadronic modes propagate in a fixed effective gravitational background which encodes salient properties of the QCD dual theory, such
as the ultraviolet conformal limit at the AdS boundary at $z \to 0$, as well as modifications of the background geometry for large $z$ which yields
confinement.  The modified theory generates the point-like hard behavior expected from QCD, instead of the soft
behavior characteristic of extended objects.~\cite{Polchinski:2001tt}  It is striking that the QCD dimensional counting rules~\cite{Brodsky:1973kr, Matveev:ra} are also a key feature of nonperturbative  models~\cite{Polchinski:2001tt} based on the gauge/gravity duality.  Although the mechanisms are different, both the perturbative QCD and the  AdS/QCD approaches depend on the leading-twist (dimension minus spin) interpolating operators of the hadrons and their structure at short distances.

Incorporating the AdS/CFT correspondence as a useful guide,  light-front holographic methods  were originally introduced~\cite{Brodsky:2006uqa}
by matching the expression for electromagnetic (EM) current matrix
elements in AdS space~\cite{Polchinski:2002jw} with the corresponding matrix elements  using light-front  theory in physical space
time.  This allows us to identify the holographic variable $z$ in AdS space with an impact variable $\zeta$.~\cite{Brodsky:2006uqa}
It was also shown that one
obtains  identical holographic mapping using the matrix elements
of the gravitational or energy-momentum tensor~\cite{Brodsky:2008pf} by perturbing the AdS metric~\cite{Abidin:2008ku}
around its static solution (\ref{eq:AdSz}).

Light-front quantization is the ideal framework to describe the
structure of hadrons in terms of their quark and gluon degrees of
freedom. The simple structure of the light-front (LF) vacuum allows an unambiguous
definition of the partonic content of a hadron in QCD and of hadronic light-front wavefunctions (LFWFs),
the underlying link between large distance hadronic states and the
constituent degrees of freedom at short distances. The LFWFs of relativistic bound states in QCD provide a description of the structure and internal dynamics of hadronic states in terms of their constituent quark and gluons  at the same LF time  $\tau = x^0 + x^3$, the time marked by the
front of a light wave,~\cite{Dirac:1949cp} instead of the ordinary instant time $t = x^0$. The constituent spin and orbital angular momentum properties of the hadrons are also encoded in the LFWFs.  In fact, the definition of quark and gluon angular momentum is unambiguous in Dirac's
front form in light-cone gauge $A^+=0$, and the gluons  have physical polarization $S^z_g= \pm 1$.
Unlike ordinary instant-time quantization, the Hamiltonian equation of motion in the light-front is frame independent and has a structure similar to  eigenmode equations in AdS space.
This makes a direct connection of QCD with AdS/CFT methods possible. The identification of orbital angular momentum of the constituents is a key element in our description of the internal structure of hadrons using holographic principles,
since hadrons with the same quark content, but different orbital angular momentum, have different masses.
In the usual AdS/QCD approach~\cite{Erlich:2005qh, DaRold:2005zs}  fields in the bulk geometry are introduced to match the
chiral symmetries of QCD and axial and vector currents, and these become the
primary entities as in effective chiral theory. In contrast, in light-front holography a direct connection with the internal constituent structure of
hadrons is established using light-front quantization.~\cite{Brodsky:2006uqa,  Brodsky:2008pf, deTeramond:2008ht}

A physical hadron in four-dimensional Minkowski space has four-momentum $P_\mu$ and invariant
hadronic mass states determined by the light-front
Lorentz-invariant Hamiltonian equation for the relativistic bound-state system
$P_\mu P^\mu \vert  \psi(P) \rangle = \mathcal{M}^2 \vert  \psi(P) \rangle$,
where the operator $P_\mu P^\mu$ is determined canonically from the QCD Lagrangian.
The physical  states in AdS space are
represented by normalizable modes $\Phi_P(x, z) = e^{-iP \cdot x} \Phi(z)$,
with plane waves along Minkowski coordinates $x^\mu$ and a profile function $\Phi(z)$
along the holographic coordinate $z$. The hadronic invariant mass
$P_\mu P^\mu = \mathcal{M}^2$  is found by solving the eigenvalue problem for the
AdS wave equation. Each  light-front hadronic state $\vert \psi(P) \rangle$ is dual to a normalizable string mode $\Phi_P(x,z)$.
For fields near the AdS boundary $z \to 0$, the behavior of $\Phi(z)$
depends on the scaling (twist) dimension of the corresponding interpolating operators.

We have shown recently a remarkable
connection between the description of hadronic modes in AdS space and
the Hamiltonian formulation of QCD in physical space-time quantized
on the light-front at equal light-front time  $\tau$.~\cite{deTeramond:2008ht}
The light-front Hamiltonian equation of motion in physical space-time
 are  the relativistic wave equations which determine the off-shell dynamics of the frame-independent light-front bound-state wavefunctions.~\cite{deTeramond:2008ht}
 To a first semiclassical approximation, where quantum loops and quark masses are not included, this leads to a light-front Hamiltonian equation which describes the bound-state dynamics of light hadrons  in terms of an invariant impact variable $\zeta$, which measures the separation of the partons within the hadron at equal light-front time.
In fact, this procedure leads to relativistic  light-front wave equations which are equivalent to the equations of motion which describe the propagation of hadronic modes in AdS space.~\cite{deTeramond:2008ht}
Remarkably, the AdS equations correspond to the kinetic energy terms of  the partons inside a hadron, whereas the interaction terms build confinement and
correspond to the truncation of AdS space in an effective dual gravity  approximation.~\cite{deTeramond:2008ht}
Early attempts to derive effective one-body equations in light-front QCD are described in reference.~\cite{Pauli:2002tj}
We should also mention previous work by 't Hooft, who obtained
the spectrum of two-dimensional QCD in the large $N_C$ limit
in terms of a Schr\"odinger equation as a function of the parton $x$-variable.~\cite{'tHooft:1974hx}
In the scale-invariant limit, this equation is equivalent
to the equation of motion for a scalar field in AdS$_3$ space.~\cite{Katz:2007br} In this case,
there is a mapping between  the variable $x$ and the radial coordinate in AdS$_3$.

\section{Higher Spin Hadronic Modes in AdS Space \label{HS} \footnote{This section is based on our collaboration with Hans Guenter Dosch and Josh Erlich. Further details will be given in a joint paper~\cite{BDDE:2010xx} and an upcoming Physics Report.}}

The description of higher spin modes in AdS space is a notoriously difficult problem.~\cite{Fronsdal:1978vb, Fradkin:1986qy, Metsaev:2008fs}
A spin-$J$ field in AdS$_{d+1}$ is represented by a rank $J$ tensor field $\Phi(x^M)_{M_1 \cdots M_J}$, which is totally symmetric in all its indices. Such a tensor contains also lower spins, which can be eliminated by imposing gauge conditions. The action for a spin-$J$ field in AdS$_{d+1}$ space time in presence of a dilaton background field $\varphi(z)$ is given by
\begin{eqnarray} \label{SJ} \nonumber
S = \half \int \! d^d x \, dz  \,\sqrt{g} \,e^{\varphi(z)}
  \Big( g^{N N'} g^{M_1 M'_1} \cdots g^{M_J M'_J} D_N \Phi_{M_1 \cdots M_J} D_{N'}  \Phi_{M'_1 \cdots M'_J}  \\
 - \mu^2  g^{M_1 M'_1} \cdots g^{M_J M'_J} \Phi_{M_1 \cdots M_J} \Phi_{M'_1 \cdots M'_J}  + \cdots \Big)  ,
\end{eqnarray}
where $D_M$ is the covariant derivative which includes parallel transport
\begin{equation} \label{Dco}
[D_N, D_K]  \Phi_{M_1 \cdots M_J} =  - R^L_{\, M_1 N K} \Phi_{L \cdots M_J} - \cdots  - R^L_{\, M_J N K} \Phi_{M_1 \cdots L},
\end{equation}
and the omitted terms refer to
terms with different contractions.  Conformal invariance in (\ref{SJ}) is broken by  the dilaton profile $\varphi(z)$ which is a function of the holographic coordinate $z$ and vanishes
 in the conformal limit $z \to 0$.  The coordinates of AdS are the Minkowski coordinates $x^\mu$ and the holographic variable $z$ labeled $x^M = \left(x^\mu, z\right)$,
 with $M, N = 1, \cdots , d+1$.

 A physical hadron has plane-wave solutions and polarization indices $\mu_i$, $i = 1 \cdots J$, along the 3 + 1 physical coordinates
 $\Phi_P(x,z)_{\mu_1 \cdots \mu_J} = e^{- i P \cdot x} \Phi(z)_{\mu_1 \cdots \mu_J}$,
 with four-momentum $P_\mu$ and  invariant hadronic mass  $P_\mu P^\mu \! = \! \mathcal{M}^2$. All other components vanish identically:
 $\Phi_{z \mu_2 \cdots \mu_J} = \cdots = \Phi_{\mu_ 1 \mu_2 \cdots z} = 0$. One can then construct an effective action in terms
 of high spin modes $\Phi_{\mu_1 \mu_2 \cdots \mu_J}$, with only the physical degrees of freedom.~\cite{Karch:2006pv, BDDE:2010xx} In this case the system of coupled differential equations which follow from (\ref{SJ}) reduce to a homogeneous equation in terms of the field   $\Phi_P(x,z)_{\mu_1 \cdots \mu_J}$ with all the polarization indices along the physical 3 + 1 coordinates.
In terms of  fields with tangent indices
\begin{equation}
 {\hat \Phi}_{A_1 A_2 \cdots A_J}
 = e_{A_1}^{M_1} e_{A_2}^{M_2} \cdots e_{A_J}^{M_J} \,
 {\Phi}_{M_1 M_2 \cdots M_J}
 =  \left(\frac{z}{R}\right)^J  \negthinspace {\Phi}_{A_1 A_2 \cdots A_J}  ,
 \end{equation}
we find the effective action~\cite{BDDE:2010xx}
\begin{eqnarray}
S = \half \int \! d^d x \, dz  \,\sqrt{g} \,e^{\varphi(z)}
  \Big( g^{N N'} \eta^{\mu_1 \mu'_1} \cdots \eta^{\mu_J \mu'_J} \partial_N \hat \Phi_{\mu_1 \cdots \mu_J} \partial_{N'} \hat  \Phi_{\mu'_1 \cdots \mu'_J}  \\ \nonumber
 - \mu^2  \eta^{\mu_1 \mu'_1} \cdots \eta^{\mu_J \mu'_J} \hat \Phi_{\mu_1 \cdots \mu_J} \hat  \Phi_{\mu'_1 \cdots \mu'_J}  \Big)  ,
\end{eqnarray}
upon rescaling the fifth dimensional mass $\mu$. The vielbein $e^A_M$
 is defined by $g_{M N} = e^A_M e^B_N\eta_{A B}$,
 where  $A, B = 1, \cdots , d+1$ are tangent AdS space indices.

The variation of the fifth-dimensional action gives the AdS wave equation  for the spin-$J$ mode $\Phi_{\mu_1 \cdots \mu_J}$
\begin{equation} \label{eq:AdSWEJ}
\left[-\frac{ z^{d-1 -2 J}}{e^{\varphi(z)}}   \partial_z \left(\frac{e^\varphi(z)}{z^{d-1 - 2 J}} \partial_z\right)
+ \left(\frac{\mu R}{z}\right)^2\right] \Phi(z)_{\mu_1 \cdots \mu_J} = \mathcal{M}^2 \Phi(z)_{\mu_1 \cdots \mu_J},
 \end{equation}
 where  $\hat \Phi(z)_{\mu_1 \cdots \mu_J} = (z/R)^{J}  \Phi(z)_{\mu_1 \cdots \mu_J}$ has scaling  behavior $\hat \Phi_J(z \to 0) \sim z^\tau$
 and scaling dimension $\tau$ given by  the relation $(\mu R)^2 = (\tau - J)(\tau -  d + J)$.
Eq. (\ref{eq:AdSWEJ}) is the basic equation which we shall use here  to describe  the propagation of spin-$J$   hadronic modes in a fixed gravitational background, asymptotic to AdS space. Its eigenvalues determine the invariant mass $\mathcal{M}^2$ of hadronic states $\vert \psi(P)\rangle$ dual to the hadronic modes $\Phi_P(x,z)_{\mu_1 \cdots \mu_J}$. We describe below how Eq. (\ref{eq:AdSWEJ})  is mapped  to the QCD light-front wave equations in physical space-time. A detailed derivation of (\ref{eq:AdSWEJ}) and comparison with the scaling described in  Ref. ~\cite{Karch:2006pv} will be given
in Ref. \cite{BDDE:2010xx}. The wave equation (\ref{eq:AdSWEJ}) also follows from considering the wave equation for a scalar mode $\Phi$ in AdS and rescaling the solution by shifting dimensions according to $\Phi_J(z) = ( z/R)^{-J}  \Phi(z)$.~\cite{deTeramond:2008ht, deTeramond:2010ge}

\section{Light-Front Quantization of QCD\label{LFquant}}

One can express the  hadron four-momentum  generator $P =  (P^+, P^-, \mbf{P}_{\!\perp})$,
$P^\pm = P^0 \pm P^3$,  in terms of the
dynamical fields, the Dirac field $\psi_+$, where $\psi_\pm = \Lambda_\pm
\psi$, $\Lambda_\pm = \gamma^0 \gamma^\pm$, and the transverse field
$\mbf{A}_\perp$ in the $A^+ = 0$ gauge~\cite{Brodsky:1997de}
quantized on the light-front at fixed light-cone time $x^+ $, $x^\pm = x^0 \pm x^3$
\begin{eqnarray} \label{eq:Pm}
P^-  &\!\!=\!&  \half \int \! dx^- d^2 \mbf{x}_\perp \bar \psi_+ \, \gamma^+
\frac{ \left( i \mbf{\nabla}_{\! \perp} \right)^2 + m^2 }{ i \partial^+}  \psi_+
 + {\rm(interactions)} ,\\ \label{eq:Pp}
P^+ &\!\!=\!& \int \! dx^- d^2 \mbf{x}_\perp
 \bar \psi_+ \gamma^+   i \partial^+ \psi_+ ,  \\ \label{eq:Pperp}
 \mbf{P}_{\! \perp}  &\!\!=\!&  \half \int \! dx^- d^2 \mbf{x}_\perp
 \bar \psi_+ \gamma^+   i \mbf{\nabla}_{\! \perp} \psi_+   ,
\end{eqnarray}
where the integrals are over the null plane $x^+ = 0$, the hyper-plane tangent to the light cone. This is the initial-value surface  for the fields where the
commutation relations are fixed.
The LF Hamiltonian $P^-$ generates LF time translations
$\left[\psi_+(x), P^-\right] = i \frac{\partial}{\partial x^+} \psi_+(x)$, to evolve the initial conditions to all space-time,
whereas the LF longitudinal  $P^+$ and  transverse momentum $\mbf{P}_\perp$ are kinematical generators. For simplicity we have omitted from (\ref{eq:Pm} - \ref{eq:Pperp}) the contributions from the gluon field $\mbf{A}_\perp$.

According to Dirac's classification of the forms of relativistic dynamics,~\cite{Dirac:1949cp} the fundamental generators of the Poincar\'e group
can be separated into kinematical and dynamical generators. The kinematical generators   act along the initial surface and leave the light-front plane invariant: they are thus independent of the dynamics and therefore contain no interactions. The dynamical generators change the light-front position and depend consequently
on  the interactions.   In addition to $P^+$ and $\mbf{P}_\perp$, the kinematical generators in the light-front frame are the $z$-component of the angular momentum $J^z$ and the boost operator $\mbf{K}$.  In addition to the Hamiltonian $P^-$,  $J^z$ and $J^y$ are also  dynamical generators. The light-front frame has the maximal number of kinematical generators.

A physical hadron in four-dimensional Minkowski space has four-momentum $P_\mu$ and invariant
hadronic mass states $P_\mu P^\mu = \mathcal{M}^2$ determined by the
Lorentz-invariant Hamiltonian equation for the relativistic bound-state system
\begin{equation} \label{eq:LFH}
P_\mu P^\mu \vert  \psi(P) \rangle = \left( P^- P^+ -  \mbf{P}_\perp^2\right) \vert \psi(P) \rangle =
\mathcal{M}^2 \vert  \psi(P) \rangle,
\end{equation}
where  the hadronic state $\vert\psi\rangle$ is an expansion in multiparticle Fock eigenstates
$\vert n \rangle$ of the free light-front  Hamiltonian:
$\vert \psi \rangle = \sum_n \psi_n \vert \psi \rangle$.
The  state $\vert \psi(P^+,\mbf{P}_\perp,J^z) \bigr\rangle$
is an eigenstate of the total momentum $P^+$
and $\mbf{P}_{\! \perp}$ and the  total spin  $J^z$. Quark and gluons appear from the light-front quantization
of the excitations of the dynamical fields $\psi_+$ and $\mbf{A}_\perp$, expanded in terms of creation and
annihilation operators at fixed LF time $\tau$. The Fock components $\psi_n(x_i, {\mathbf{k}_{\perp i}}, \lambda_i^z)$
are independent of  $P^+$ and $\mbf{P}_{\! \perp}$
and depend only on relative partonic coordinates:
the momentum fraction
 $x_i = k^+_i/P^+$, the transverse momentum  ${\mathbf{k}_{\perp i}}$ and spin
 component $\lambda_i^z$. Momentum conservation requires
 $\sum_{i=1}^n x_i = 1$ and
 $\sum_{i=1}^n \mathbf{k}_{\perp i}=0$.
The LFWFs $\psi_n$ provide a
{\it frame-independent } representation of a hadron which relates its quark
and gluon degrees of freedom to their asymptotic hadronic state.

\section{A Semiclassical Approximation to QCD \label{LFholog}}

We can compute $\mathcal{M}^2$ from the hadronic matrix element
\begin{equation}
\langle \psi(P') \vert P_\mu P^\mu \vert\psi(P) \rangle  =
\mathcal{M}^2  \langle \psi(P' ) \vert\psi(P) \rangle,
\end{equation}
expanding the initial and final hadronic states in terms of its Fock components. The computation is  simplified in the
frame $P = \big(P^+, \mathcal{M}^2/P^+, \vec{0}_\perp \big)$ where $P^2 =  P^+ P^-$.
We find
 \begin{equation} \label{eq:Mk}
\mathcal{M}^2  =  \sum_n  \! \int \! \big[d x_i\big]  \! \left[d^2 \mbf{k}_{\perp i}\right]
 \sum_q \Big(\frac{ \mbf{k}_{\perp q}^2 + m_q^2}{x_q} \Big)
 \left\vert \psi_n (x_i, \mbf{k}_{\perp i}) \right \vert^2  + {\rm (interactions)} ,
 \end{equation}
plus similar terms for antiquarks and gluons ($m_g = 0)$. The integrals in (\ref{eq:Mk}) are over
the internal coordinates of the $n$ constituents for each Fock state
\begin{equation}
\int \big[d x_i\big] \equiv
\prod_{i=1}^n \int dx_i \,\delta \Bigl(1 - \sum_{j=1}^n x_j\Bigr) , ~~~
\int \left[d^2 \mbf{k}_{\perp i}\right] \equiv \prod_{i=1}^n \int
\frac{d^2 \mbf{k}_{\perp i}}{2 (2\pi)^3} \, 16 \pi^3 \,
\delta^{(2)} \negthinspace\Bigl(\sum_{j=1}^n\mbf{k}_{\perp j}\Bigr),
\end{equation}
with phase space normalization
$\sum_n  \int \big[d x_i\big] \left[d^2 \mbf{k}_{\perp i}\right]
\,\left\vert \psi_n(x_i, \mbf{k}_{\perp i}) \right\vert^2 = 1$.

The LFWF $\psi_n(x_i, \mathbf{k}_{\perp i})$ can be expanded in terms of  $n-1$ independent
position coordinates $\mathbf{b}_{\perp j}$,  $j = 1,2,\dots,n-1$,
conjugate to the relative coordinates $\mbf{k}_{\perp i}$, with $\sum_{i = 1}^n \mbf{b}_{\perp i} = 0$.
We can also express (\ref{eq:Mk})
in terms of the internal impact coordinates $\mbf{b}_{\perp j}$ with the result
\begin{equation}
 \mathcal{M}^2  =  \sum_n  \prod_{j=1}^{n-1} \int d x_j \, d^2 \mbf{b}_{\perp j} \,
\psi_n^*(x_j, \mbf{b}_{\perp j})  \\
 \sum_q   \left(\frac{ \mbf{- \nabla}_{ \mbf{b}_{\perp q}}^2  \! + m_q^2 }{x_q} \right)
 \psi_n(x_j, \mbf{b}_{\perp j}) \\
  + {\rm (interactions)} . \label{eq:Mb}
 \end{equation}
The normalization is defined by
$\sum_n  \prod_{j=1}^{n-1} \int d x_j d^2 \mathbf{b}_{\perp j}
\left \vert \psi_n(x_j, \mathbf{b}_{\perp j})\right\vert^2 = 1$.
To simplify the discussion we will consider a two-parton hadronic bound state.  In the limit
of zero quark mass
$m_q \to 0$
\begin{equation}  \label{eq:Mbpion}
\mathcal{M}^2  =  \int_0^1 \! \frac{d x}{x(1-x)} \int  \! d^2 \mbf{b}_\perp  \,
  \psi^*(x, \mbf{b}_\perp)
  \left( - \mbf{\nabla}_{ {\mbf{b}}_{\perp}}^2\right)
  \psi(x, \mbf{b}_\perp) +   {\rm (interactions)}.
 \end{equation}

The functional dependence  for a given Fock state is
given in terms of the invariant mass
\begin{equation}
 \mathcal{M}_n^2  = \Big( \sum_{a=1}^n k_a^\mu\Big)^2 = \sum_a \frac{\mbf{k}_{\perp a}^2 +  m_a^2}{x_a}
 \to \frac{\mbf{k}_\perp^2}{x(1-x)} \,,
 \end{equation}
giving  the measure of  the off-energy shell of the bound state,
 $\mathcal{M}^2 \! - \mathcal{M}_n^2$.
 Similarly, in impact space the relevant variable for a two-parton state is  $\zeta^2= x(1-x)\mbf{b}_\perp^2$.
Thus, to first approximation  LF dynamics  depend only on the boost invariant variable
$\mathcal{M}_n$ or $\zeta$,
and hadronic properties are encoded in the hadronic mode $\phi(\zeta)$ from the relation
\begin{equation} \label{eq:psiphi}
\psi(x,\zeta, \varphi) = e^{i L \varphi} X(x) \frac{\phi(\zeta)}{\sqrt{2 \pi \zeta}} ,
\end{equation}
thus factoring out the angular dependence $\varphi$ and the longitudinal, $X(x)$, and transverse mode $\phi(\zeta)$.
This is a natural factorization in the light front since the
corresponding canonical generators, the longitudinal and transverse generators $P^+$ and $\mbf{P}_\perp$ and the $z$-component of the orbital angular momentum
$J^z$ are kinematical generators which commute with the LF Hamiltonian generator $P^-$.
We choose the normalization of the LF mode $\phi(\zeta) = \langle \zeta\vert \psi\rangle$ as $\langle \phi \vert \phi \rangle =
\int d\zeta \, \vert \langle  \zeta \vert \phi \rangle \vert^2 = 1$.

We can write the Laplacian operator in (\ref{eq:Mbpion}) in circular cylindrical coordinates $(\zeta, \varphi)$
and factor out the angular dependence of the
modes in terms of the $SO(2)$ Casimir representation $L^2$ of orbital angular momentum in the
transverse plane. Using  (\ref{eq:psiphi}) we find~\cite{deTeramond:2008ht}
\begin{equation} \label{eq:KV}
\mathcal{M}^2   =  \int \! d\zeta \, \phi^*(\zeta) \sqrt{\zeta}
\left( -\frac{d^2}{d\zeta^2} -\frac{1}{\zeta} \frac{d}{d\zeta}
+ \frac{L^2}{\zeta^2}\right)
\frac{\phi(\zeta)}{\sqrt{\zeta}}   \\
+ \int \! d\zeta \, \phi^*(\zeta) U(\zeta) \phi(\zeta) ,
\end{equation}
where $L = L^z$.  In writing the above equation we have summed the complexity of the interaction terms in the QCD Lagrangian  by the introduction of the effective
potential $U(\zeta)$, which is  modeled to enforce confinement at some IR scale.
The LF eigenvalue equation $P_\mu P^\mu \vert \phi \rangle  =  \mathcal{M}^2 \vert \phi \rangle$
is thus a light-front  wave equation for $\phi$
\begin{equation} \label{LFWE}
\left(-\frac{d^2}{d\zeta^2}
- \frac{1 - 4L^2}{4\zeta^2} + U(\zeta) \right)
\phi(\zeta) = \mathcal{M}^2 \phi(\zeta),
\end{equation}
a relativistic single-variable LF Schr\"odinger equation.   Its eigenmodes $\phi(\zeta)$
determine the hadronic mass spectrum and represent the probability
amplitude to find $n$-partons at transverse impact separation $\zeta$,
the invariant separation between pointlike constituents within the hadron~\cite{Brodsky:2006uqa} at equal LF time.
Extension of the results to arbitrary $n$ follows from the $x$-weighted definition of the
transverse impact variable of the $n-1$ spectator system~\cite{Brodsky:2006uqa}:
$\zeta = \sqrt{\frac{x}{1-x}} \left\vert \sum_{j=1}^{n-1} x_j \mbf{b}_{\perp j} \right\vert$, where $x = x_n$ is the longitudinal
momentum fraction of the active quark. One can also
generalize the equations to allow for the kinetic energy of massive
quarks using Eqs. (\ref{eq:Mk}) or (\ref{eq:Mb}). In this case, however,
the longitudinal mode $X(x)$ does not decouple from the effective LF bound-state equations.

\section{Light-Front Holographic Mapping and Hadronic Spectrum}

The structure of the QCD Hamiltonian equation  (\ref{eq:LFH}) for the state $\vert \psi(P) \rangle$ is similar to the structure of the wave equation (\ref{eq:AdSWEJ}) for the
$J$-mode $\Phi_{\mu_1 \cdots \mu_J}$ in AdS; they are both frame-independent and have identical eigenvalues $\mathcal{M}^2$, the mass spectrum of the color-singlet states of QCD. This provides a profound connection between physical QCD and the physics of hadronic modes in AdS space. However, important differences are also apparent:  Eq. (\ref{eq:LFH}) is a linear quantum-mechanical equation of states in Hilbert space, whereas Eq. (\ref{eq:AdSWEJ}) is a classical gravity equation; its solutions describe spin-$J$ modes propagating in a higher dimensional
warped space. Physical hadrons are composite and thus inexorably endowed of orbital angular momentum. Thus, the identification
of orbital angular momentum is of primary interest in establishing a connection between both approaches.

As shown in  Sect. \ref{LFholog}, one can indeed systematically reduce  the LF  Hamiltonian eigenvalue Eq.  (\ref{eq:LFH}) to an effective relativistic wave equation (\ref{LFWE}), analogous to the AdS equations, by observing that each $n$-particle Fock state has an essential dependence on the invariant mass of the system  and
thus, to a first approximation, LF dynamics depend only on $\mathcal{M}_n^2$.
In  impact space the relevant variable is the boost invariant  variable $\zeta$,
 which measures the separation of the  constituents and which also allows one to separate the dynamics
of quark and gluon binding from the kinematics of the constituent
internal angular momentum.

Upon the substitution $z \! \to\! \zeta$  and
$\phi_J(\zeta)   = \left(\zeta/R\right)^{-3/2 + J} e^{\varphi(z)/2} \, \Phi_J(\zeta)$,
in (\ref{eq:AdSWEJ}), we find for $d=4$ the QCD light-front wave equation (\ref{LFWE}) with effective potential~\cite{deTeramond:2010ge}
\begin{equation} \label{U}
U(\zeta) = \half \varphi''(z) +\frac{1}{4} \varphi'(z)^2  + \frac{2J - 3}{2 z} \varphi'(z) .
\end{equation}
The fifth dimensional mass $\mu$ is not a free parameter but scales as $(\mu R)^2 = - (2-J)^2 + L^2$.

If $L^2 < 0$, the LF Hamiltonian  is unbounded from below
$\langle \phi \vert P_\mu P^\mu \vert \phi \rangle <0$  and the spectrum contains an
infinite number of unphysical negative values of $\mathcal{M}^2 $ which can be arbitrarily large.
As $\mathcal{M}^2$ increases in absolute value, the particle becomes
localized within a very small region near $\zeta = 0$, since  the effective potential is conformal at small  $\zeta$.
For $\mathcal{M}^2 \to - \infty$ the particle is localized at $\zeta = 0$, the
particle ``falls towards the center''.~\cite{LL:1958}
The critical value  $L=0$  corresponds to the lowest possible stable solution, the ground state of the light-front Hamiltonian.
For $J = 0$ the five dimensional mass $\mu$
 is related to the orbital  momentum of the hadronic bound state by
 $(\mu R)^2 = - 4 + L^2$ and thus  $(\mu R)^2 \ge - 4$. The quantum mechanical stability condition $L^2 \ge 0$ is thus equivalent to the
 Breitenlohner-Freedman stability bound in AdS.~\cite{Breitenlohner:1982jf}
The scaling dimensions are $2 + L$ independent of $J$, in agreement with the
twist-scaling dimension of a two-parton bound state in QCD.
It is important to notice that in the light-front the $SO(2)$ Casimir for orbital angular momentum $L^2$
is a kinematical quantity, in contrast with the usual $SO(3)$ Casimir $L(L+1)$ from non-relativistic physics which is
rotational, but not boost invariant. 

We consider here the  positive-sign dilaton profile $\exp(+ \kappa^2 z^2)$ which explicitly confines the constituents  to distances
$\langle z \rangle \sim 1/\kappa$~\cite {deTeramond:2009xk, Andreev:2006ct}. It is also the solution compatible with (\ref{eq:AdSWEJ}),
which naturally encodes the internal structure of hadrons and their orbital angular momentum.
From (\ref{U}) we obtain  the effective potential~\cite{deTeramond:2009xk}
$U(\zeta) =   \kappa^4 \zeta^2 + 2 \kappa^2(L + S - 1)$,  where $J^z = L^z + S^z$, which  corresponds  to a transverse oscillator in the light-front.
Equation  (\ref{LFWE}) has eigenfunctions
\begin{equation} \label{phi}
\phi_{n, L}(\zeta) = \kappa^{1+L} \sqrt{\frac{2 n!}{(n\!+\!L\!)!}} \, \zeta^{1/2+L}
e^{- \kappa^2 \zeta^2/2} L^L_n(\kappa^2 \zeta^2) ,
\end{equation}
and eigenvalues
\begin{equation} \label{M2}
\mathcal{M}_{n, L, S}^2 = 4 \kappa^2 \left(n + L + \frac{S}{2} \right).
\end{equation}

The meson spectrum  has a string-theory Regge form: the square of the masses are linear with the same slope in both the internal orbital angular momentum $L$ and radial quantum number $n$, where $n$ counts the number of nodes  of the wavefunction in the radial variable $\zeta$. The spectrum also depends on the internal spin $S$.
The lowest possible solution for $n = L = S = 0$ has eigenvalue $\mathcal{M}^2 = 0$.
This is a chiral symmetric bound state of two massless quarks with scaling dimension 2 and size
 $\langle \zeta^2 \rangle \sim 1/\kappa^2$, which we identify with the lowest state, the pion.
Thus one can compute the hadron spectrum by simply adding  $4 \kappa^2$ for a unit change in the radial quantum number, $4 \kappa^2$ for a change in one unit in the orbital quantum number and $2 \kappa^2$ for a change of one unit of spin to the ground state value of $\mathcal{M}^2$. Remarkably, the same rule holds for
baryons.~\cite{deTeramond:2010we}  This is an important feature of light-front holography, which predicts the same multiplicity of states for mesons
and baryons  as it is observed experimentally.~\cite{Klempt:2007cp}

\begin{figure}[h]
\begin{center}
\includegraphics[width=6.4cm]{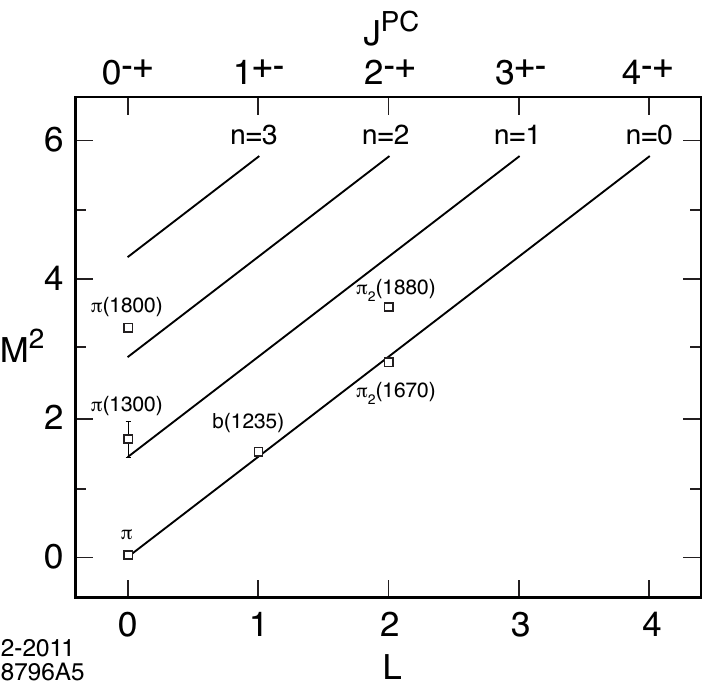}  \hspace{10pt}
\includegraphics[width=6.4cm]{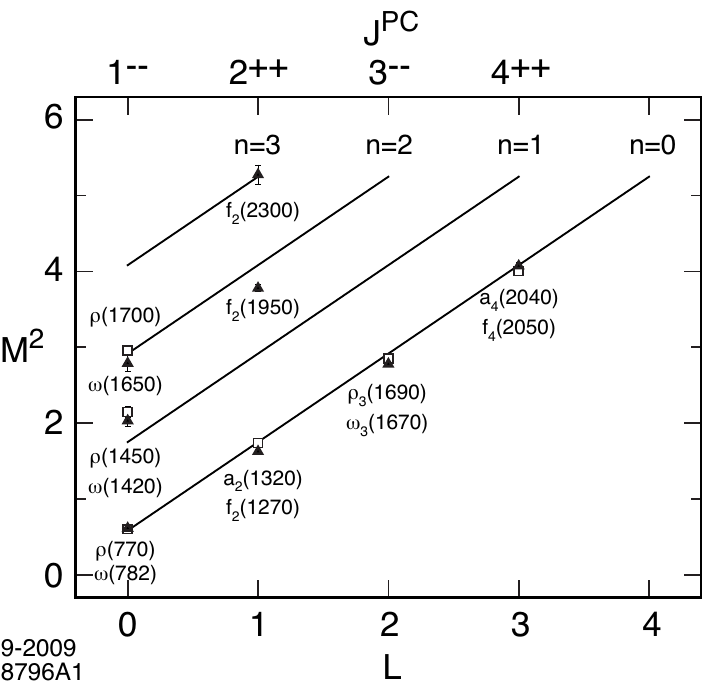}
 \caption{\small Regge trajectories for   $\pi$-meson family with
$\kappa= 0.6$ GeV (left);    $I\!=\!1$ $\rho$-meson
 and $I\!=\!0$  $\omega$-meson families with $\kappa= 0.54$ GeV (right). Only confirmed PDG states~\cite{Amsler:2008xx} are shown.}
\label{pionspec}
\end{center}
\end{figure}

Individual hadron states are identified by their interpolating operators at $z\to 0.$ Pion interpolating operators are constructed by examining the behavior of
bilinear covariants $\bar \psi \Gamma \psi$ under charge conjugation and parity transformation.
Thus, for example, a pion interpolating operator $\bar q \gamma_5 q$ create a state with quantum numbers $J^{PC} = 0^{- +}$, and a vector meson
interpolating operator $\bar q \gamma_\mu q$ a state $1^{- -}$. Likewise the operator $\bar q \gamma_\mu \gamma_5 q$ creates a state with
$1^{++}$ quantum numbers,  the $a_1(1260)$ positive parity meson. If we include  orbital excitations the pion interpolating operator is
$\mathcal{O}_{2+L} = \bar q \gamma_5  D_{\{\ell_1} \cdots D_{\ell_m\}} q$. This is an operator  with total internal  orbital
momentum $L = \sum_{i=1}^m \ell_i$, twist $\tau = 2 + L$ and canonical dimension $\Delta = 3 + L$.  The scaling of  the AdS field $\Phi(z) \sim z^\tau$ at $z \to 0$  is precisely the scaling required to match the scaling dimension of the local meson interpolating operators.   The spectral predictions for  light meson and vector meson  states are compared with experimental data
in Fig. \ref{pionspec} for the positive sign dilaton model discussed here.

\begin{figure}[h]
\begin{center}
\includegraphics[angle=0,width=12.0cm]{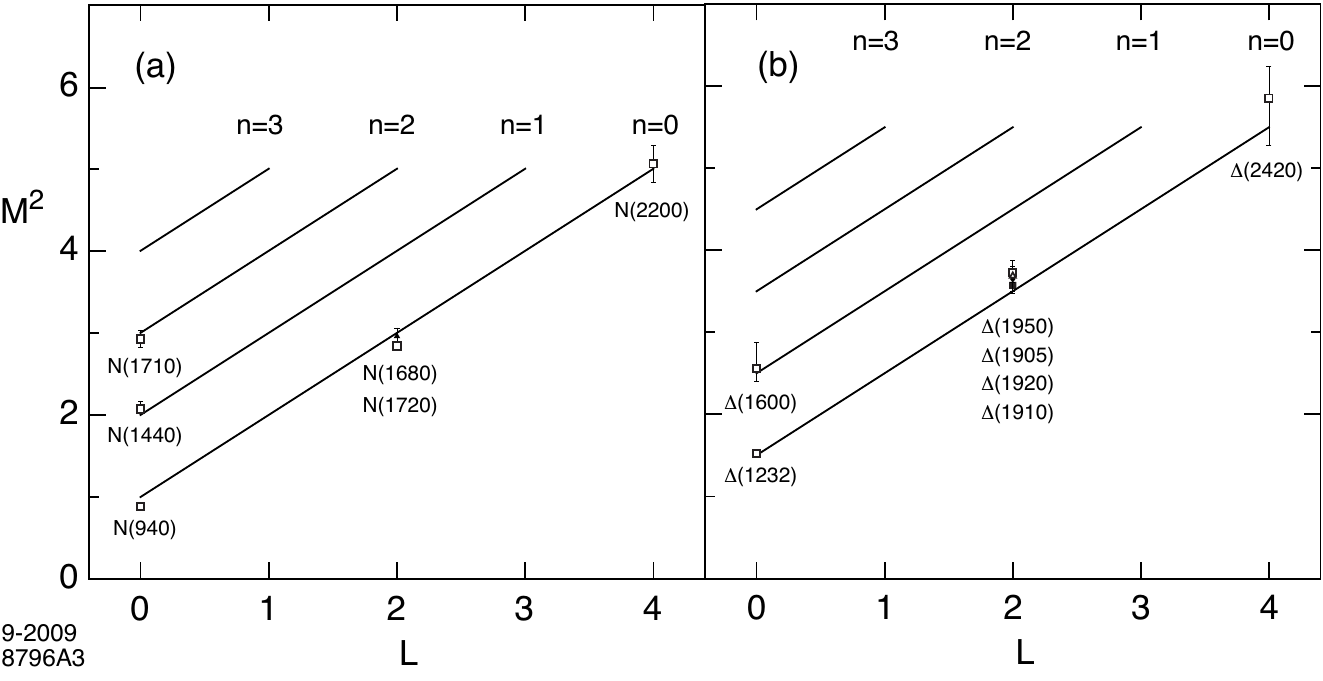}  
\caption{\small Positive parity  Regge trajectories for  the  $N$ and $\Delta$
baryon families for $\kappa= 0.5$ GeV.
Data from \cite{Amsler:2008xx}. }
\label{baryons}
\end{center}
\end{figure}

The predictions for the positive-parity light baryons  are shown in Fig. \ref{baryons}. As in the meson sector,  the increase  in the
mass squared for  higher baryonic states is
$\Delta n = 4 \kappa^2$, $\Delta L = 4 \kappa^2$ and $\Delta S = 2 \kappa^2$,
relative to the lowest ground state,  the proton.~\cite{deTeramond:2010we} Only confirmed
PDG~\cite{Amsler:2008xx} states are shown.
The Roper state $N(1440)$ and the $N(1710)$ are well accounted for in this model as the first  and second radial
states. Likewise, the $\Delta(1660)$ corresponds to the first radial state of the $\Delta$ family. The model is  successful in explaining the important parity degeneracy observed in the light baryon spectrum, such as the $L\! =\!2$, $N(1680)\!-\!N(1720)$ pair and the $\Delta(1905), \Delta(1910), \Delta(1920), \Delta(1950)$ states which are degenerate
within error bars. The parity degeneracy of baryons is also a property of the ``hard wall" model, but radial states are not well described by this
model.~\cite{deTeramond:2005su}
For other
recent calculations of the hadronic spectrum in the framework of AdS/QCD, see Refs.~\cite{BoschiFilho:2005yh, Evans:2006ea, Hong:2006ta, Colangelo:2007pt, Forkel:2007ru, Vega:2008af, Nawa:2008xr,  dePaula:2008fp, Colangelo:2008us, Forkel:2008un, Ahn:2009px, Sui:2009xe, Kapusta:2010mf, Zhang:2010bn, Iatrakis:2010zf, Branz:2010ub, Kirchbach:2010dm, Sui:2010ay}.

The proton in light-front holography is described by  a two component wave function
$\psi(\zeta) = \psi_+(\zeta) u_+  + \psi_-(\zeta)u_- $,
 where $u_\pm$ are four-dimensional spinors.
The  $L^z=0$ and $L^z= + 1$ orbital components are combined with spin components $S^z = + 1/2$ and $S^z = -1/2$ respectively.
An interesting feature of light-front holography for baryons and massless quarks is the fact that the lowest valence Fock states with $L^z=0$  and $L^z=\pm 1$ have the same probability $ \int d \zeta\,  \vert \psi_+(\zeta) \vert^2 = \int d \zeta \, \vert \psi_-(\zeta) \vert^2$,
 a manifestation of the  chiral invariance of the theory for massless quarks.~\cite{Brodsky:2010px}
 This implies that the quarks carry zero angular momentum $\langle S_z=0\rangle$ in the proton with $J^z=\pm 1/2$ and $\langle L^z=1/2\rangle$.

\section{Light Front Holographic Mapping of Current Matrix Elements}

The great advantage of the front form  -- as emphasized by Dirac --  is that boost operators are kinematic.  Unlike the instant form,  the boost operators in the front form have no interaction terms.  The calculation of a current matrix element $\langle P + q \vert J^\mu \vert P \rangle$ requires boosting the hadronic  eigenstate from $\vert P \rangle $ to $\vert P + q \rangle $, a task which becomes hopelessly complicated in the instant form.

The form factor is computed in the light front from the matrix elements of the plus-component of the current $J^+$, in order to avoid coupling to Fock states with different numbers of constituents
\begin{equation}
\langle \psi(P') \vert J^+(0) \vert \psi(P) \rangle = (P+P')^+ F(q^2),
\end{equation}
 where ~$Q = P' \! - P$ ~and~  $J^+(x) \! = \sum_q e_q(x) \bar q \gamma^+ q(x)$ is the quark current which couples locally to the quarks in the hadron. Expanding the  initial and final Fock states in terms of Fock components, we obtain
 Drell-Yan-West (DYW) expression~\cite{Drell:1969km, West:1970av} upon integration over the intermediate variables in the $q^+$ frame:
 \begin{equation} \label{eq:DYW}
F(q^2) = \sum_n  \int \big[d x_i\big] \left[d^2 \mbf{k}_{\perp i}\right]
\sum_j e_j \psi^*_n (x_i, \mbf{k}'_{\perp i},\lambda_i)
\psi_n (x_i, \mbf{k}_{\perp i},\lambda_i),
\end{equation}
where the variables of the light cone Fock components in the
final-state are given by $\mbf{k}'_{\perp i} = \mbf{k}_{\perp i}
+ (1 - x_i)\, \mbf{q}_\perp $ for a struck  constituent quark and
$\mbf{k}'_{\perp i} = \mbf{k}_{\perp i} - x_i \, \mbf{q}_\perp$ for each
spectator. The formula is exact if the sum is over all Fock states $n$.
The form factor can also be conveniently written in impact space
\begin{equation} \label{eq:FFb}
F(q^2) =  \sum_n  \prod_{j=1}^{n-1}\int d x_j d^2 \mbf{b}_{\perp j}
\exp \! {\Bigl(i \mbf{q}_\perp \! \cdot \sum_{j=1}^{n-1} x_j \mbf{b}_{\perp j}\Bigr)}
\left\vert  \psi_n(x_j, \mbf{b}_{\perp j})\right\vert^2,
\end{equation}
corresponding to a change of transverse momentum $x_j \mbf{q}_\perp$ for each
of the $n-1$ spectators.

On the higher dimensional gravity theory, the hadronic matrix element  corresponds to
the  non-local coupling of an external electromagnetic field $A^M(x,z)$  propagating in AdS with the extended mode $\Phi(x,z)$~\cite{Polchinski:2002jw}
 \begin{equation}
 \int d^4x~dz~\sqrt{g} ~  e^{\varphi(z)} A^{M}(x,z)
 \Phi^*_{P'}(x,z) \overleftrightarrow\partial_M \Phi_P(x,z).
 \end{equation}
Can  the transition amplitudes be related? How can we recover hard pointlike scattering at large $Q$ from the soft collision of extended objects?~\cite{Polchinski:2002jw}
Although the expressions for the transition amplitudes look very different, one can show that a precise mapping of the $J^+$ elements  can be carried out at fixed light-front time, using the semiclassical approximation discussed in Sect. \ref{LFholog}. In fact, if both expressions for the form factor are identical for arbitrary values of $q$, then the factorization  (\ref{eq:psiphi}) is required. For a two
parton state  we find $\phi(\zeta)   = \left(\zeta/R\right)^{-3/2} e^{\varphi(z)/2} \, \Phi(\zeta)$ as before, and $X(x) = \sqrt{x(1-x)}$. In this case the longitudinal mode does not decouple, but instead a specific form for $X(x)$ is obtained.~\cite {Brodsky:2006uqa}  A precise mapping can be established for an
arbitrary number of partons.~\cite {Brodsky:2006uqa} Identical results are obtained for the mapping of the matrix elements of the energy momentum
tensor.~\cite{Brodsky:2008pf}

\begin{figure}[h]  \hspace{1.2cm}
\includegraphics[angle=0,width=6.6cm]{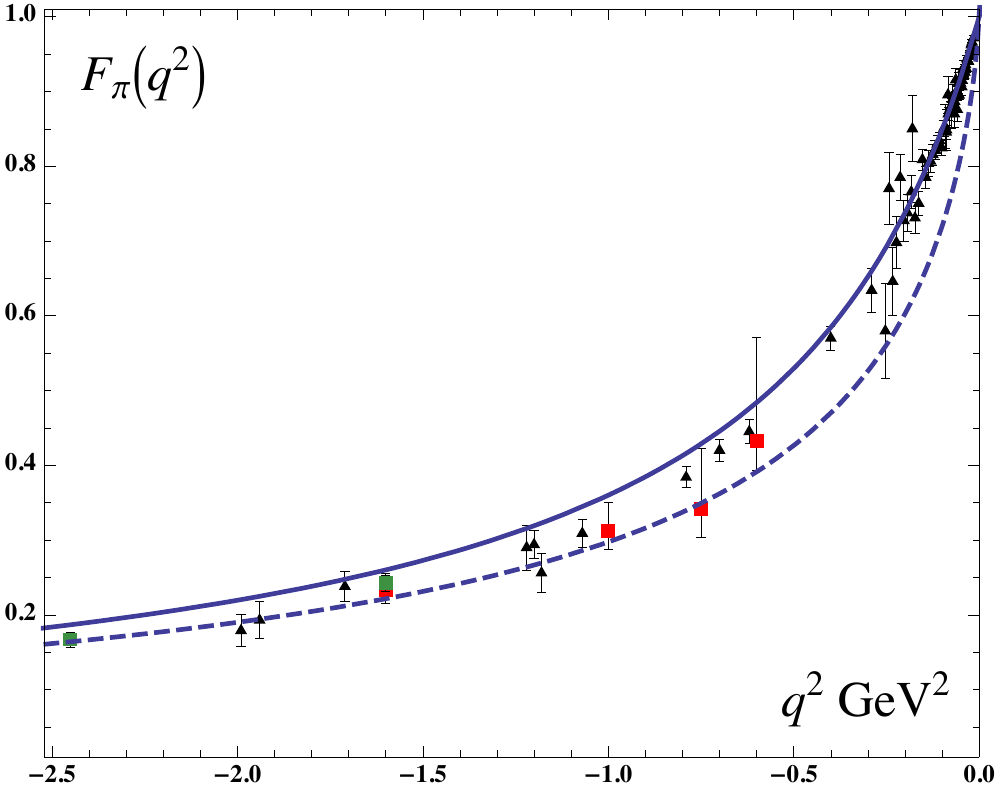} \hspace{0.8cm} 
\begin{minipage}[b]{14pc}
\caption{\label{PionFF} \small Space-like pion form factor $F_\pi(q^2)$.  Continuous line: confined current, dashed  line free current.
Triangles are the data compilation  from Baldini~\cite{Baldini:1998qn},  boxes  are JLAB data.~\cite{Tadevosyan:2007yd}}
\end{minipage}
\end{figure}

The results described above correspond to a ``free" current propagating on AdS space and dual to the EM pointlike current in the DYW LF formula, which allow us to map state-by-state.~\footnote{In general the mapping relates the AdS density $\Phi^2(z)$ to an  effective LF single particle transverse density.~\cite {Brodsky:2006uqa}}
This mapping has the shortcoming that the pole structure of the form factor is not built on the timelike region. Furthermore, the moments of the form factor at $Q^2=0$ diverge, giving for example an infinite charge radius.  The pole structure is generated when the EM current is confined, this means, when the EM current propagates on
a IR deformed AdS space to mimic confinement. This also leads to finite moments at $Q^2=0$, as illustrated on Fig. \ref{PionFF}.  Since the computation of the form factor involves a twist-3 current $J^+$, the poles do not correspond to the physical poles of the twist-2 transverse current $\mbf{J}_\perp$ present in the annihilation channel. Consequently, the location of the poles in the final result should be rescaled to their physical positions. When this is done the results agree extremely well with  available data. The non-perturbative effects from the ``dressed" current correspond to an infinite sum of diagrams. One should however be careful to avoid a double counting of terms. This important point deserves further investigation.

\section{Conclusions}

The light-front holographic approach described in this paper provides a direct correspondence between an effective gravity theory defined in a fifth-dimensional warped space with a  semiclassical approximation to strongly coupled QCD quantized on the light-front. This duality leads to a  Lorentz-invariant relativistic Schr\"odinger wave equation~\cite{deTeramond:2008ht} which
provides a successful prediction for the light-quark meson and baryon spectra as
a function of hadron spin, quark angular momentum, and radial quantum numbers. It also predicts the same multiplicity of states for mesons
and baryons which is observed experimentally. Thus this AdS/QCD approach encodes the salient features of QCD.

We originally derived light-front holography using the identity between electromagnetic and gravitational form factors computed in AdS and light-front theory.~\cite{Brodsky:2006uqa, Brodsky:2008pf}
The results for hadronic form factors are also successful, and the predicted power law fall-off agrees with dimensional counting rules as required by conformal invariance at small $z$. In particular, we have shown the importance of including  non-perturbative effects corresponding to the infinite sum of diagrams contained in the  structure of the currents propagating in the confining geometry. This ``dressed" current is required to reproduce the time-like pole structure and finite moments near $Q^2 =0$, effects which are not seen in the semiclassical approximation when comparing state by state.
The semiclassical approximation to light-front QCD described in this paper does not account for particle creation and absorption; it is thus expected to break down at short distances where hard gluon exchange and quantum corrections become important. Other shortcomings of holographic methods are discussed in Ref.~\cite{Csaki:2008dt}.

\vspace{10pt}

\noindent {\bf Acknowledgements}

\noindent Invited talk presented by GdT at  XIV School of Particles and Fields,  Morelia, M\'exico, November 8-12,  2010. GdT is grateful to the organizers for their outstanding hospitality. We are very grateful to  H. G. Dosch and J. Erlich for their helpful comments and collaboration. This research was supported by the Department
of Energy  contract DE--AC02--76SF00515. SLAC-PUB-14393.

\end{document}